\documentstyle[pre,preprint,aps]{revtex}

\begin{document}

%
%
%
%
%
\catcode`\@=11\relax
\newwrite\@unused
\def\typeout#1{{\let\protect\string\immediate\write\@unused{#1}}}
\typeout{psfig: version 1.1}
\def\psglobal#1{
\typeout{psfig: including #1 globally}
\immediate\special{ps:plotfile #1 global}}
\def\psfiginit{\typeout{psfiginit}
\immediate\psglobal{/usr/lib/ps/figtex.pro}}
%
%
\def\@nnil{\@nil}
\def\@empty{}
\def\@psdonoop#1\@@#2#3{}
\def\@psdo#1:=#2\do#3{\edef\@psdotmp{#2}\ifx\@psdotmp\@empty \else
    \expandafter\@psdoloop#2,\@nil,\@nil\@@#1{#3}\fi}
\def\@psdoloop#1,#2,#3\@@#4#5{\def#4{#1}\ifx #4\@nnil \else
       #5\def#4{#2}\ifx #4\@nnil \else#5\@ipsdoloop #3\@@#4{#5}\fi\fi}
\def\@ipsdoloop#1,#2\@@#3#4{\def#3{#1}\ifx #3\@nnil 
       \let\@nextwhile=\@psdonoop \else
      #4\relax\let\@nextwhile=\@ipsdoloop\fi\@nextwhile#2\@@#3{#4}}
\def\@tpsdo#1:=#2\do#3{\xdef\@psdotmp{#2}\ifx\@psdotmp\@empty \else
    \@tpsdoloop#2\@nil\@nil\@@#1{#3}\fi}
\def\@tpsdoloop#1#2\@@#3#4{\def#3{#1}\ifx #3\@nnil 
       \let\@nextwhile=\@psdonoop \else
      #4\relax\let\@nextwhile=\@tpsdoloop\fi\@nextwhile#2\@@#3{#4}}
\def\psdraft{
	\def\@psdraft{0}
}
\def\psfull{
	\def\@psdraft{100}
}
\psfull
\newif\if@prologfile
\newif\if@postlogfile
\newif\if@bbllx
\newif\if@bblly
\newif\if@bburx
\newif\if@bbury
\newif\if@height
\newif\if@width
\newif\if@rheight
\newif\if@rwidth
\newif\if@clip
\def\@p@@sclip#1{\@cliptrue}
\def\@p@@sfile#1{
		   \def\@p@sfile{#1}
}
\def\@p@@sfigure#1{\def\@p@sfile{#1}}
\def\@p@@sbbllx#1{
		\@bbllxtrue
		\dimen100=#1
		\edef\@p@sbbllx{\number\dimen100}
}
\def\@p@@sbblly#1{
		\@bbllytrue
		\dimen100=#1
		\edef\@p@sbblly{\number\dimen100}
}
\def\@p@@sbburx#1{
		\@bburxtrue
		\dimen100=#1
		\edef\@p@sbburx{\number\dimen100}
}
\def\@p@@sbbury#1{
		\@bburytrue
		\dimen100=#1
		\edef\@p@sbbury{\number\dimen100}
}
\def\@p@@sheight#1{
		\@heighttrue
		\dimen100=#1
   		\edef\@p@sheight{\number\dimen100}
}
\def\@p@@swidth#1{
		\@widthtrue
		\dimen100=#1
		\edef\@p@swidth{\number\dimen100}
}
\def\@p@@srheight#1{
		\@rheighttrue
		\dimen100=#1
		\edef\@p@srheight{\number\dimen100}
}
\def\@p@@srwidth#1{
		\@rwidthtrue
		\dimen100=#1
		\edef\@p@srwidth{\number\dimen100}
}
\def\@p@@sprolog#1{\@prologfiletrue\def\@prologfileval{#1}}
\def\@p@@spostlog#1{\@postlogfiletrue\def\@postlogfileval{#1}}
\def\@cs@name#1{\csname #1\endcsname}
\def\@setparms#1=#2,{\@cs@name{@p@@s#1}{#2}}
%
%
\def\ps@init@parms{
		\@bbllxfalse \@bbllyfalse
		\@bburxfalse \@bburyfalse
		\@heightfalse \@widthfalse
		\@rheightfalse \@rwidthfalse
		\def\@p@sbbllx{}\def\@p@sbblly{}
		\def\@p@sbburx{}\def\@p@sbbury{}
		\def\@p@sheight{}\def\@p@swidth{}
		\def\@p@srheight{}\def\@p@srwidth{}
		\def\@p@sfile{}
		\def\@p@scost{10}
		\def\@sc{}
		\@prologfilefalse
		\@postlogfilefalse
		\@clipfalse
}
%
%
\def\parse@ps@parms#1{
	 	\@psdo\@psfiga:=#1\do
		   {\expandafter\@setparms\@psfiga,}}
%
%
\newif\ifno@bb
\newif\ifnot@eof
\newread\ps@stream
\def\bb@missing{
	\typeout{psfig: searching \@p@sfile \space  for bounding box}
	\openin\ps@stream=\@p@sfile
	\no@bbtrue
	\not@eoftrue
	\catcode`\%=12
	\loop
		\read\ps@stream to \line@in
		\global\toks200=\expandafter{\line@in}
		\ifeof\ps@stream \not@eoffalse \fi
		\@bbtest{\toks200}
		\if@bbmatch\not@eoffalse\expandafter\bb@cull\the\toks200\fi
	\ifnot@eof \repeat
	\catcode`\%=14
}	
\catcode`\%=12
\newif\if@bbmatch
\def\@bbtest#1{\expandafter\@a@\the#1
\long\def\@a@#1
\long\def\bb@cull#1 #2 #3 #4 #5 {
	\dimen100=#2 bp\edef\@p@sbbllx{\number\dimen100}
	\dimen100=#3 bp\edef\@p@sbblly{\number\dimen100}
	\dimen100=#4 bp\edef\@p@sbburx{\number\dimen100}
	\dimen100=#5 bp\edef\@p@sbbury{\number\dimen100}
	\no@bbfalse
}
\catcode`\%=14
\def\compute@bb{
		\no@bbfalse
		\if@bbllx \else \no@bbtrue \fi
		\if@bblly \else \no@bbtrue \fi
		\if@bburx \else \no@bbtrue \fi
		\if@bbury \else \no@bbtrue \fi
		\ifno@bb \bb@missing \fi
		\ifno@bb \typeout{FATAL ERROR: no bb supplied or found}
			\no-bb-error
		\fi
		\count203=\@p@sbburx
		\count204=\@p@sbbury
		\advance\count203 by -\@p@sbbllx
		\advance\count204 by -\@p@sbblly
		\edef\@bbw{\number\count203}
		\edef\@bbh{\number\count204}
}
%
%
\def\in@hundreds#1#2#3{\count240=#2 \count241=#3
		     \count100=\count240	
		     \divide\count100 by \count241
		     \count101=\count100
		     \multiply\count101 by \count241
		     \advance\count240 by -\count101
		     \multiply\count240 by 10
		     \count101=\count240	
		     \divide\count101 by \count241
		     \count102=\count101
		     \multiply\count102 by \count241
		     \advance\count240 by -\count102
		     \multiply\count240 by 10
		     \count102=\count240	
		     \divide\count102 by \count241
		     \count200=#1\count205=0
		     \count201=\count200
			\multiply\count201 by \count100
		 	\advance\count205 by \count201
		     \count201=\count200
			\divide\count201 by 10
			\multiply\count201 by \count101
			\advance\count205 by \count201
		     \count201=\count200
			\divide\count201 by 100
			\multiply\count201 by \count102
			\advance\count205 by \count201
		     \edef\@result{\number\count205}
}
\def\compute@wfromh{
		\in@hundreds{\@p@sheight}{\@bbw}{\@bbh}
		\edef\@p@swidth{\@result}
}
\def\compute@hfromw{
		\in@hundreds{\@p@swidth}{\@bbh}{\@bbw}
		\edef\@p@sheight{\@result}
}
\def\compute@handw{
		\if@height 
			\if@width
			\else
				\compute@wfromh
			\fi
		\else 
			\if@width
				\compute@hfromw
			\else
				\edef\@p@sheight{\@bbh}
				\edef\@p@swidth{\@bbw}
			\fi
		\fi
}
\def\compute@resv{
		\if@rheight \else \edef\@p@srheight{\@p@sheight} \fi
		\if@rwidth \else \edef\@p@srwidth{\@p@swidth} \fi
}
%
\def\compute@sizes{
	\compute@bb
	\compute@handw
	\compute@resv
}
%
%
\def\psfig#1{\vbox {
	%
	\ps@init@parms
	\parse@ps@parms{#1}
	\compute@sizes
	\ifnum\@p@scost<\@psdraft{
		\typeout{psfig: including \@p@sfile \space }
		\special{ps::[begin] 	\@p@swidth \space \@p@sheight \space
				\@p@sbbllx \space \@p@sbblly \space
				\@p@sbburx \space \@p@sbbury \space
				startTexFig \space }
		\if@clip{
			\typeout{(clip)}
			\special{ps:: \@p@sbbllx \space \@p@sbblly \space
				\@p@sbburx \space \@p@sbbury \space
				doclip \space }
		}\fi
		\if@prologfile
		    \special{ps: plotfile \@prologfileval \space } \fi
		\special{ps: plotfile \@p@sfile \space }
		\if@postlogfile
		    \special{ps: plotfile \@postlogfileval \space } \fi
		\special{ps::[end] endTexFig \space }
		\vbox to \@p@srheight true sp{
			\hbox to \@p@srwidth true sp{
				\hfil
			}
		\vfil
		}
	}\else{
		\vbox to \@p@srheight true sp{
		\vss
			\hbox to \@p@srwidth true sp{
				\hss
				\@p@sfile
				\hss
			}
		\vss
		}
	}\fi
}}
\catcode`\@=12\relax

\pagestyle{empty}
\narrowtext
\tighten
\textwidth= 6.5in
\textheight= 9.5in

\centerline{\bf MICROSCOPIC ESTIMATES FOR ELECTROMIGRATION VELOCITIES OF}
\centerline{\bf INTRAGRANULAR VOIDS IN THIN ALUMINUM LINES}
\noindent
\\
L. K. WICKHAM, James P. SETHNA \\
Laboratory of Atomic and Solid--State Physics,
Cornell University,
Ithaca,  NY 14853--2501

\bigskip
\noindent
{\bf ABSTRACT}
\bigskip

We explore the effect of faceting on possible mechanisms for mass transport
around electromigration voids in aluminum interconnects.
Motivated by linear
response estimates which suggest that particle flux would be much higher
along steps than across terraces on a clean aluminum surface, we study
step nucleation in the presence of a small driving force along a surface.
We find that step nucleation, even on a nearly defect-free void surface,
 would be slow if the step energy is equal to that calculated for a clean
aluminum surface.
In the presence of a uniform electromigration force,
the creation of new steps between existing ones should not occur unless
the free energy cost of a step is much less than thermal energies.
We conclude that voids cannot move intragranularly at $\mu$m/hr rates
without help from other factors such as local heating and impurities.

\bigskip
\noindent
{\bf INTRODUCTION}

\bigskip

The increasing miniaturization of aluminum interconnects in VLSI circuits
has produced an urgent need for new insight into electromigration phenomena.
Although researchers have arrived at useful models for
particle motion along grain boundaries in wide aluminum lines, lines
with micron width or smaller
tend to have a ``bamboo'' structure which lacks continuous grain boundary
paths, and different electromigration
mechanisms can dominate in these small lines. Electromigration voids move
through single grains, typically along the edge of a line\cite{motion}.
Failure often occurs in the middle of a grain,
where voids can collide to make a large hole
or one void can collapse into a transverse slit\cite{Marieb}\cite{failure}.
Electron microscope observations
suggest that collapse tends to occur after a void moves into a grain
with the appropriate orientation\cite{slit}. Thus, it seems that void motion is
crucial to both of these forms of failure. But what microscopic processes
are actually happening as those voids move?

One great help in modeling this flow is the weakness of the electromigration
driving force. The force $F$ induced by a current density $j$ and resistivity
$\rho$ follows the linear relationship $F = Z^{\star} ~e ~\rho ~j$,
where observations of
macroscopic behavior give the proportionality constant in aluminum as high
as $Z^{\star}\sim 20$ \cite{Z}.
For typical testing conditions in aluminum interconnects, with
current density $> 10^6 A/cm^2$ and temperature 500-600K,
this driving force has to move an
atom through roughly an entire micron before it does 1 $kT$ of work. Although
current concentration near voids, resistivity variation at the void surface,
and orientation of the void surface
will modify the size of the electromigration force,
the bulk value provides a useful starting estimate for surface calculations,
and it certainly indicates that the driving force is a small perturbation to
surface diffusion.

For simplicity, previous models of flow around the edge of a void have often
assumed a cylindrical void, where all atoms on the
surface participate in diffusion with an estimated average mobility\cite{Li}.
In contrast, observed
voids in thin interconnects usually show clear facets, particularly
on their leading face. Examination shows such facets
to have (111) orientation\cite{lateMarieb}\cite{facets}. Since the
energy required to dislodge an atom in an Al(111) terrace is very high, flow
across void a surface should probably be modeled by first asking how
many adatoms (or vacancies) are available for surface diffusion, and then,
if this number is very low,
checking to see if motion along steps should dominate the mass flux.
The latter will depend, in part, on step density at the void surface, and we
will explore this issue at some length.

\bigskip
\noindent
{\bf Linear Response on Al(111)}

\bigskip

To calculate the lowest order term in this flux, we assume that the
density of the mobile species is unchanged when the force is turned on and
simply moves according to linear response:
$$
	flux ~=~ D ~(F/kT) ~n(x) ~-~ D~ dn/dx,    \eqno(1)
$$
where D is the diffusion constant and n(x) is the local density of mobile
diffusers.
Near a straight step, the equilibrium density of adatoms at a certain
type of site is given by the free energy cost of
moving an atom from a kink site on the step to the type of
spot in question, which might be on the surface or resting against the
step\cite{BCF}. Thus, we model the first term above as
a simple function of force, temperature, and diffusion barriers.
Recent density functional calculations carried out in the local density
approximation (LDA) by Stumpf and Scheffler\cite{StumpfScheffler} allow an
estimate of this lowest order flux term for adatoms on both a clean Al(111)
surface and at steps on that surface parallel to the driving force
\cite{vacancies}. To do so, we assume a typical diffusion attempt rate of
$10^{12}$ Hz and obtain flux estimates from (1) in units of
atoms/s along an atom column. (The actual attempt rate for the exchange
 process in \{111\} step diffusion may be an order of magnitude higher
\cite{StumpfScheffler}.)

\bigskip

\begin{tabular}{ccccccc}
\hline
structure&
\multicolumn{1}{c}{ $~$adatom cost  } &
\multicolumn{1}{c}{ $~~$ad. density } &
\multicolumn{1}{c}{ $~~$diffusion barrier } &
\multicolumn{1}{c}{ $~~$hop time }&
\multicolumn{1}{c}{ $~$atoms/s/column}
\\
\hline
$(111)$ surface & 1.05 eV & $8 \cdot 10^{-10}$ & $.04$ eV & $2 \cdot 10^{-12}s$
& .1\\
$\{111\}$ step&    .28 eV & $4 \cdot 10^{-3}$  & $.42$ eV & $4 \cdot 10^{-9}s$
& 300\\
$\{100\}$ step&  .25 eV & $7 \cdot 10^{-3}$ & $.32$ eV & $6 \cdot 10^{-10}s$
& 3500\\
\hline
\end{tabular}

\bigskip

\vbox{
{\bf \narrower {Table I.
Estimates at 580K and current density $\simeq 5 \cdot 10^6$ A$/$cm$^2$,
 using density functional results by Stumpf and Scheffler.}
}}
\label{tab:experiments}

\bigskip Under testing conditions,
observed void velocities seem to range from a micron every few hours
to over 1$\mu$ m/hr (although some voids are stationary).
A micron sized void with velocity
1 $\mu$ m/hr would
have to move roughly $4 \cdot 10^3$ atoms/s through each atom column
on its surface. The estimates above would only approach this flux for
a dense network of steps.
The estimates in Table I are not, of course, a quantitative prediction for
adatom flow in real electromigration voids. Stumpf and Scheffler find that
energy differences in their calculations are converged to within .06 eV, but
use of the local density approximation at a surface may introduce some
additional error. More importantly, voids do not have clean
aluminum surfaces, and we will discuss this issue below. Still, this
estimate of particle flux on clean aluminum steps and terraces
yields two conclusions.
First, unless void surfaces are covered with steps, {\bf something} must
be raising particle flux there far above the rates expected for clean Al (111).
Second, unless contaminants change the ratio of step to surface flow by
orders of magnitude, steps on a dirty aluminum surface will remain
competitive routes for mass flow. Both conclusions raise the issue ---
what step density should be expected on void surfaces?

\bigskip
\noindent
{\bf STEP NUCLEATION MODELS}
\bigskip

As a void moves, its leading side continually uncovers new sections of (111)
surface. Thus, if the leading surface of the void contains a high step
density, new steps must constantly nucleate there\cite{dislocation_dens}.
Since this surface of the void is losing mass, new steps will tend to appear
through pits in the surface which arise from fluctuations and will grow
if they become larger than a critical size. An external driving force can
affect step
density by changing this critical pit size, but we will find that the
weak force present during electromigration cannot produce rapid step
production.

First, we consider the most favorable case in which a driving force could
enhance step nucleation. If a pit appears where there is no incoming flux of
adatoms, the pit will reach critical size when it attains equilibrium with its
downstream environment, taken to be a straight step that is a distance
L from the pit. This equilibrium must
balance two effects: when the driving force F moves an atom from pit to step it
gives the atom an energy $-F \cdot L$, but taking an atom out of the pit
increases its size and increases the energy cost of its perimeter by an
amount $g/R$, where $g$ is a step free energy per unit length, and $R$ is the
radius of the pit. Thus, when this radius is larger than $g/(F \cdot L)$,
the pit will tend to lose more and more atoms to the downstream step; tiny pits
will disappear and not contribute any new steps to the system.

Even this most favorable configuration for step nucleation produces
a rather high estimate for critical pit size on an electromigrating
surface. Recall that the electromigration force does an amount of work of
order 1 kT (i.e. of order .05 eV) in moving an atom through a micron ---
a void size! Stumpf and Scheffler
find that step energies on the Al (111) surface are between .23 and .25
eV/ lattice site\cite{entropy_ignored}.
Thus, even our extreme model configuration with $L$
as large as a micron would lead to a critical pit radius between 4.5 and 5
lattice sites. The corresponding critical pit energy,
$g ~(2 \pi R_c) ~-~ \pi R_c^2 ~F L$, is between 3 and 4 eV for these estimates
and provides a substantial barrier to nucleation.

In a more realistic case, the pit which appears through thermal fluctuations
will lie between two steps. A stable pit must be in equilibrium with the
adatom density around it, in the mass flow between the upstream and the
downstream
step. To find the adatom density required for pit equilibrium, we turn to
a surface with no driving force. There, Gibbs--Thomson theory predicts that
mass will flow into a pit unless the adatom density around it is suppressed
relative to the equilibrium density near a straight pit by the factor
$ e^{-g/(R kT)}$\cite{GT}, where $g$ and $R$ were defined above. If a driving
force is present, can it kinetically suppress the adatom density between two
steps by a factor equal to the Gibbs--Thomson suppression near an equilibrium
pit?

\bigskip
\centerline{
\vbox{\psfig{figure=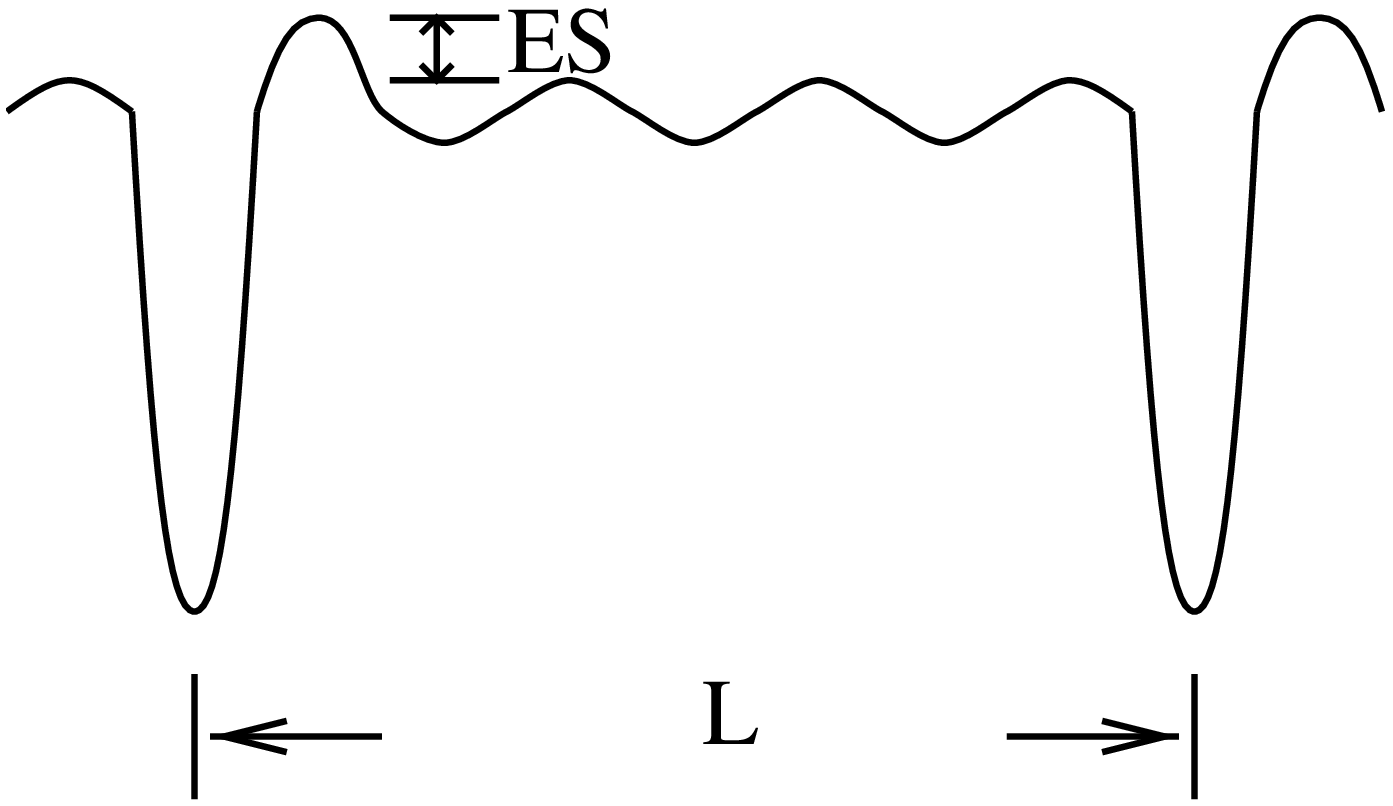,width=3.truein}}
}
\medskip {\narrower {\parskip 0pt Figure~1 Toy model of energy landscape
for adatom diffusion between steps.
}}
\bigskip

To answer this question, we consider a toy model for a parallel step array,
whose energy landscape for adatom diffusion is shown in figure 1\cite{Natori}.
The adatom density profile between the steps must be a solution of the
driven diffusion equation $D {{d^2n}/{dx^2}} - D {F\over{kT}}{dn/{dx}}=0$
. Boundary conditions must match the flux
in the central region between steps with the rate at which particles enter
and leave the supply of adatoms attached to each step. For small flux
between steps and relatively slow diffusion along a step, the density of step
adatoms should be very close to the equilibrium value\cite{arg}. Imposing this
step density as a boundary condition for the energy landscape with constant
barriers (ES=0 in figure 1), yields an adatom density between the steps which
is constant and equal to its equilibrium value. In this case, the driving
force is unable to produce the suppression of adatom density required for
a pit to be in equilibrium. Without an extra Ehrlich-Schwoebel
barrier for motion at a step, a uniform electromigration force should be
unable to induce {\bf any} pit nucleation between steps.

If there is an extra barrier for leaving a step (designated ES in figure
1), then the adatom density will vary between steps.
Figure 2 shows this
variation across a step separation of $100$ lattice sites, for $ES= 2kT$
(a value motivated by the LDA results) and
two different force strengths. These density results are relatively insensitive
to step separation. If a pit appeared between two long parallel
steps with these barriers, then far away from the
pit the adatom density would follow the solution from the toy model.
Mass would flow towards the pit from the far region unless the local adatom
density near the pit was as large as the corresponding toy model result.
Thus, a critical pit near the upstream step would have a radius such that
$e^{-g/R_c kT}$ was equal to the kinetic suppression of adatom density
near the upstream step in the toy model result. The results in figure 2
for $F a = .001$ (a large force for an electromigration problem)
show a kinetic suppression of .994, which corresponds to a huge critical
pit size unless $g/kT < 1$.
Thus, even when there is an extra Ehrlich-Schwoebel barrier at steps,
the weak driving force expected in electromigration should be quite
ineffective at inducing pit nucleation.

\bigskip
\centerline{\vbox
{\psfig{figure=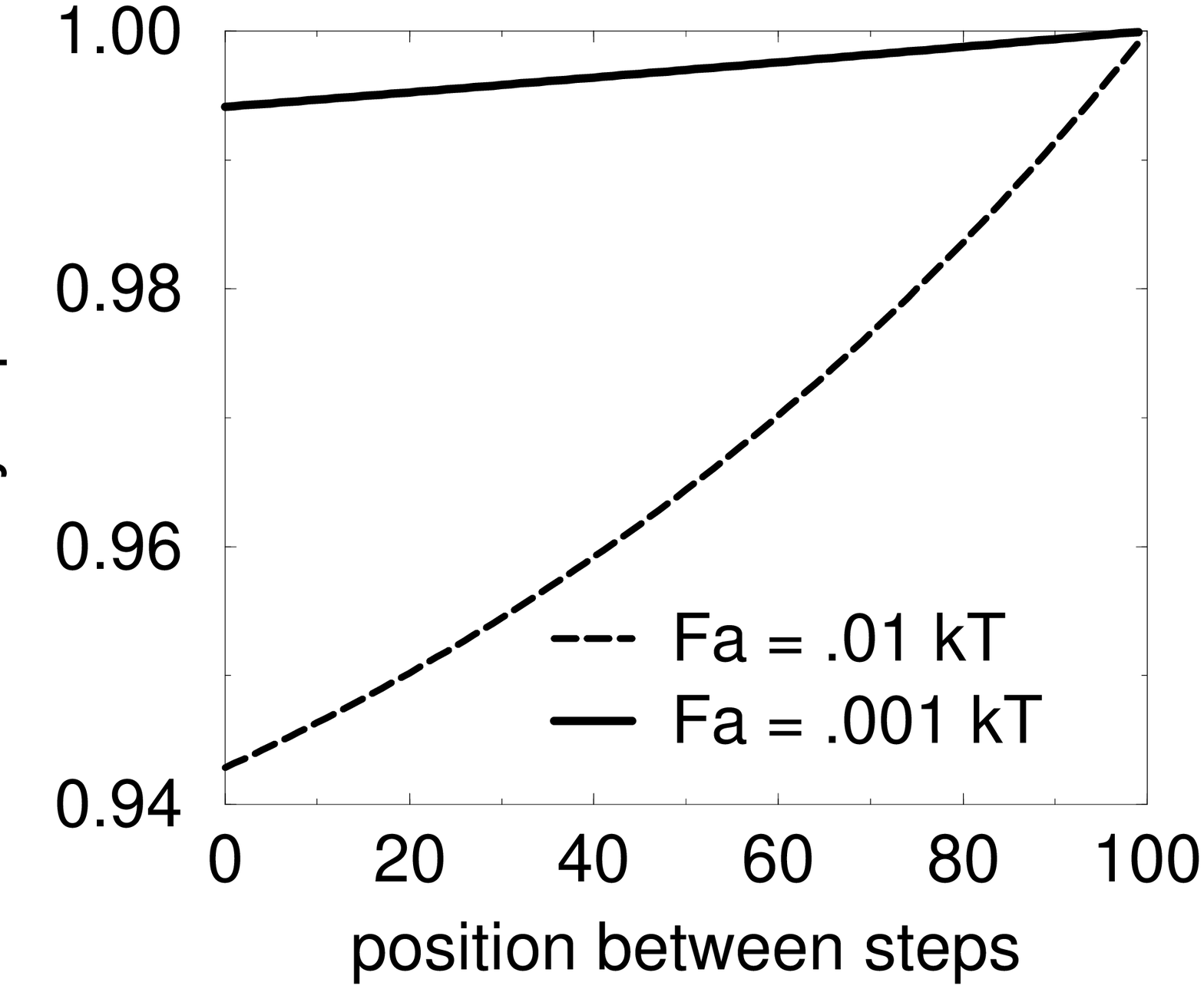,width=3.5truein}}
}

\medskip

\noindent
{Figure 2.
Adatom density between steps
as fraction of equilibrium value for $ES = 2kT$ and two values of the
driving force. Note that the smaller force value here is larger than
that in many electromigration tests and yet corresponds to a minimal
kinetic suppression of the adatom density. (Note: $a$ is a lattice
spacing.)

\bigskip
\noindent
{\bf OTHER INFLUENCES}

\bigskip

So, if an electromigration void contained ideal Al (111) surfaces, then the
leading face of the void would contain very few steps and the flow
across that surface would be far too low to allow the void to move at a
$\mu$m/hr rate at testing temperatures. Yet, voids do move.
What additional factors are important in real voids?
First, some evidence exists for intense local heating near voids at large
current densities.
Where such thermal spikes were measured, current density
throughout the wire was an astounding 48 MA/cm$^2$\cite{Kondo}. Other
observations suggested that local melting can occur near voids which
nearly breach an interconnect\cite{Marieb}. There, the macroscopic current
density of 5MA/cm$^2$ would have been magnified by current concentration.
At such huge current densities, impediments to
heat flow, such as local loss of thermal contact with the base, might
produce the dramatic heating observed. Since such heating effects go
as $J^2$, they should be far less important
at normal testing currents with less constriction. Novel heating mechanisms
may also be important near $10^8 A/cm^2$.\cite{Sorbello}

In addition, real void surfaces are not clean.
Many interconnects are doped with copper, which
is known to produce a small enhancement of bulk
aluminum diffusion\cite{copper}. If residual etchant
from the lithography process contaminates large regions of the void surface,
it could dramatically change aluminum flow and surface structure. And,
of course, exposed void surfaces will oxidize.
Recent observations of interface mass loss suggest that mobility at
a boundary between aluminum and a full aluminum oxide layer can be high ---
perhaps approaching that along the fast diffusion paths at aluminum grain
boundaries\cite{Augur}. Still, this phenomena may not be ubiquitous, since
STM observations found that oxidation of the first layer or so of an Al(111)
surface suppressed step diffusion and appeared to freeze the surface
topography in place\cite{Brune}.

Even if a specific impurity is found to be key for rapid
void velocity, a determination of the mechanism for  enhancement will
still be of interest.
If contaminants can eliminate most of the free energy cost of step production
on the close--packed Al (111) surface, then the leading surfaces of
electromigration voids may be covered with enough steps to explain
observed transport. If not, these surfaces should be highly faceted and
enhanced flow must occur between steps.
To move $4000$ atoms/s through every column on a surface at
quoted testing temperatures, linear response
estimates require the sum of adatom (or vacancy) production cost and
corresponding diffusion barrier to be about half
that given for adatoms on clean aluminum terraces. Perhaps further study
will pinpoint a contaminant that can produce such dramatic changes
throughout a void surface.

\bigskip
\noindent
{\bf CONCLUSIONS}
\bigskip

Linear response estimates of mass flux on a clean Al(111) surface
are far too low to explain observed void motion at standard testing
conditions unless void surfaces are covered with a dense network of
steps. As layers of atoms are removed from the leading surface of a void,
the electromigration
force should be too weak to induce substantial nucleation of new steps
unless the step free energy is nearly zero.
Although heating effects seem to enhance void velocity at extremely high
current density, impurities may play a vital part in
boosting void velocity at standard testing conditions.
Two different mechanisms for mobility enhancement could occur ---
reduction in step energy or an increase of mass flow between
steps. Since these various mechanisms can scale quite differently with
current density and temperature, understanding the sources of high
void mobility during laboratory electromigration tests should help
in determining whether related mechanisms are important to failure
under normal interconnect operation.



\begin{references}

\bibitem{motion} Several observations of void motion appear in the
references below. Marieb {\it et al} performed SEM {\it in situ} studies of
passivated lines and Arzt {\it et al} have studied unpassivated ones.
In addition, TEM {\it in situ} studies are described in
S.P. Riege, A. W. Hunt, and J. A. Prybyla, MRS Symp. Proc {\bf 391}, p. 249.

\bibitem{Marieb} T.N. Marieb, E. Abratowski, and J. C. Bravman in
{\it Stress-induced Phenomena in Metallization}, ed. P.S. Ho, C.-Y. Li,
and P. Totta, (A.I.P. Proc. {\bf 305}, NY, 1994) pp. 1-14.

\bibitem{failure}
J. H. Rose, Appl. Phys. Lett. {\bf 61}, 2170 (1992).

\bibitem{slit}
E. Arzt, O. Kraft, W.D. Nix, and J.E. Sanchez, J. Appl. Phys. {\bf 76},
	1563 (1994).
\bibitem{lateMarieb} T. Marieb, P. Flinn, J. C. Bravman, D. Gardner, and
	M. Madden, J. Appl. Phys. {\bf 78}, 1026 (1995).

\bibitem{Z} Z. Suo, MRS Symp. Proc. {\bf 338}, p. 379.

\bibitem{Li} C.-Y. Li, P. Borgesen, and T. D. Sullivan, Appl. Phys. Lett.
	{\bf 59}, 1464 (1991);
	P. S. Ho, J. Appl. Phys. {\bf 41}, 64 (1970);
	W.D. Nix and E. Arzt, Met. Trans. A {\bf 23A}, 2007 (1992).

\bibitem{facets} S. Shingubara, H. Kaneko, M. Saitoh, J. Appl. Phys. {\bf 69}
	 207 (1991).
	 Y.-C. Joo and C.V. Thompson, MRS Symp. Proc. {\bf 309}, p. 319.

\bibitem{BCF} W. K. Burton, N. Cabrera, and F. C. Frank, Phil. Trans.
	Roy. Soc. A {\bf 243}, 299 (1951).

\bibitem{StumpfScheffler} R. Stumpf and M. Scheffler, Phys. Rev. B {\bf 53},
	4958 (1996).

\bibitem{vacancies} In the results of Stumpf and Scheffler,
the sum of diffusion barrier plus cost to produce a
mobile carrier is even higher for vacancies than adatoms on the surface.

\bibitem{entropy_ignored} Since the roughening transition of the Al(111)
surface is near its melting temperature, we use zero temperature free energies.
In the square lattice Ising model, the zero-temperature form for the step 
free energy is
accurate to within about one percent even at $T= .5 T_c$: J.E. Avron,
H. van Beijeren, L.S. Schulman, and R.K.P. Zia, J. Phys. A {\bf 15},
L81 (1982).

\bibitem{dislocation_dens} Screw dislocations can, of course,
	produce steps, but dislocation densities in interconnects have
	been estimated at $10-50 /\mu m^2$: J. H. Rose, J. R. Loyd, A.
	Shepela, and N. Riel,{\it Proc. 49th Mtng. of Elec. Microsc.
	Soc. Am.}, ed. G. W. Bailey and E. L. Hall (S. F. Press, San
	Fransisco, 1991) pp. 820-821.

\bibitem{Natori} The following step flow calculation is in a similar spirit to
that of A. Natori, H. Fujimura, and M. Fukuda, Appl. Surf. Sci {\bf 60/61},
85 (1992).

\bibitem{GT} See, for example,
	P. Wynblatt and N. A. Gjostein, in {\it Progress in Solid
	State Chemistry}, ed. J. O. McCaldin, and F. Somorjai (Pergamon,
	Oxford, 1975) vol. 9, p. 21. Tiny voids or high adatom density
	may require some correction to Gibbs-Thomson theory: see B.
	Krishnamachari, J. McLean, B. Cooper, and J. Sethna, sub. to
	Phys. Rev. B.

\bibitem{arg} Even if there is no flux entering a step, setting a
local linear
response flux leaving the step equal to $D_{step} \bigtriangledown^2
n_{step}$, the divergence at the step, gives a step adatom profile whose
variation from the equilibrium density goes as $exp(-y/\lambda )$, where
$y$ is the distance to the nearest kink and $\lambda=
\sqrt{ {D_{step} /[D_{surf}~(F/kT)~
exp(-{\rm step~detach~cost}/kT)~]}}$, roughly a thousand lattice
sites for clean aluminum parameters.


\bibitem{Kondo} S. Kondo and K. Hinode, Appl. Phys. Lett. {\bf 67},
	1606 (1995).

\bibitem{Sorbello} R.S. Sorbello, to appear in {\it Adv. Metal.
for Future ULSI}, (MRS Symp. Proc., 1996).

\bibitem{copper} A. S. Oates, J. Appl. Phys. {\bf 79}, 163 (1996).

\bibitem{oxide} M. LeGall, B. Lesage, Phil. Mag. A {\bf 70}, 761 (1994).


\bibitem{Augur} R.A. Augur, R. Wolters, W. Schmidt, and S. Kordic,
MRS Symp. Proc. {\bf 391}, p. 259.

\bibitem{1970} R. K. Hart and J. K. Maurin, Surf. Sci. {\bf 20}, 285 (1970).

\bibitem{Brune} H. Brune, J. Wintterlin, J. Trost, and G. Ertl, J. Chem.
Phys. {\bf 99}, 2128 (1993).

\end{references}
\end{document}